\documentclass[pre,twocolumn,longbibliography]{revtex4-1}
\usepackage{amsmath}
\usepackage{amssymb}
\usepackage{longtable}
\usepackage{array}
\usepackage{natbib}
\usepackage{subfigure}
\usepackage{mathtools}
\usepackage{graphicx}
\usepackage{color}
\usepackage{setspace}
\usepackage{mathrsfs}
\usepackage[usenames,dvipsnames]{xcolor}
\newcolumntype{P}[1]{>{\centering\arraybackslash}p{#1}}

\def\bdot{\raise.2em\hbox to .15em{.}}

\definecolor{gray}{gray}{0.5}

\def\bdotblack{\raise.25em\hbox to .15em{.}}

\definecolor{pinegreen}{rgb}{0.0, 0.47, 0.44}
\definecolor{debianred}{rgb}{0.84, 0.04, 0.33}

\begin{document}
\title{Recurrent flow patterns as a basis for turbulence: predicting statistics from structures}

\author{Jacob Page$^{1}$}
\author{Peter Norgaard$^{2}$}
\author{Michael P. Brenner$^{2,3}$}
\author{Rich R. Kerswell$^4$}
\affiliation{$^{1}$School of Mathematics, University of Edinburgh, Edinburgh, EH9 3FD, UK}
\affiliation{$^{2}$Google Research, Mountain View, CA 94043}
\affiliation{$^{3}$School of Engineering and Applied Sciences, Harvard University, Cambridge MA 02138}
\affiliation{$^{4}$DAMTP, Centre for Mathematical Sciences, University of Cambridge, Cambridge, CB3 0WA, UK}

\begin{abstract}

\noindent
A dynamical systems approach to turbulence envisions the flow as a trajectory through a high-dimensional state space transiently visiting the neighbourhoods of unstable simple invariant solutions (E. Hopf, {\em Commun. Appl. Maths} {\bf 1}, 303, 1948). The hope has always been to turn this appealing picture into a predictive framework where the statistics of the flow follows from a weighted sum of the statistics of each simple invariant solution. Two outstanding obstacles have prevented this goal from being achieved: $(1)$ paucity of known solutions and $(2)$ the lack of a rational theory for predicting the required weights. Here we describe a method to substantially solve these problems, and thereby  provide the first compelling evidence that the PDFs of a fully developed turbulent flow can be reconstructed with a set of unstable periodic orbits. Our new method for finding solutions uses automatic differentiation, with high-quality guesses constructed by minimising a trajectory-dependent loss function. We use this approach to find hundreds of new solutions in turbulent, two-dimensional Kolmogorov flow. Robust statistical predictions are then computed by {\sl learning} weights after converting a turbulent trajectory into a Markov chain for which the states are individual solutions, and the nearest solution to a given snapshot is determined using a deep convolutional autoencoder. To our knowledge, this is the first time the PDFs of a spatio-temporally-chaotic system have been successfully reproduced with a set of simple invariant states, and provides a fascinating connection between self-sustaining dynamical processes and the more well-known statistical properties of turbulence. 
\end{abstract}

\maketitle

\section{Introduction}

A compelling view of turbulence, originally advocated by \citet{Hopf1948}, is to consider a turbulent flow as an orbit in a very high-dimensional state space pinballing between unstable simple invariant solutions. 
This viewpoint is attractive for both mechanistic understanding, which can be found in the dynamics of the individual solutions, and for quantifying the relationship of individual dynamical events to long-time statistics. 
In recent decades attempts to realise this approach have dramatically improved our understanding of transitional shear flows \citep{Kawahara2012, Graham2021}: 
As examples, the onset of turbulence in pipes has been connected to the emergence of finite-amplitude travelling wave solutions beyond some critical Reynolds number, $Re$, in saddle node bifurcations \citep{Kerswell2005,Eckhardt2007},
while the later discovery of unstable periodic orbits (UPOs) in so-called `minimal' turbulent configurations \citep{Kawahara2001,Cvitanovic2010} has revealed some of the self-sustaining mechanisms at play in wall bounded turbulence \citep{Hamilton1995,Waleffe1997,Hall2010}.
However, extending these ideas to high-$Re$ flows has been restricted by computational limitations both in the methods finding UPOs and using them to label realisations of fully developed turbulence.

For fully developed turbulence the dynamical systems view imagines a state space littered with simple invariant solutions (equilibria, travelling waves, UPOs) whose entangled stable and unstable manifolds create the scaffold for the turbulent pinball \cite{Hopf1948,Eckhardt2002, Kerswell2005, KerswellTutty2007, Gibson2008, Cvitanovic2010, Chandler2013, Suri2020, Yalniz2021, Krygier2021}.  
This picture suggests a predictive theory based upon a suitably weighted sum of the properties of the visited invariant solutions where the weights reflect the relative time spent in the neighbourhood of the solution \citep{ChaosBook}. 
However, finding enough of the important solutions let alone identifying the appropriate weights has proved prohibitively expensive. 
This is especially so at high $Re$ where the set of invariant solutions proliferates dramatically and each becomes increasingly unstable, hampering their identification. 

A central and well-known issue is the sensitivity of the Newton-Raphson root-finding algorithm to the quality of the initial guess in high dimensional problems.  Past work has mainly relied on `recurrent flow analysis' to generate good enough guesses for UPOs  where a turbulent orbit is required to shadow a UPO for at least one full period \citep{Kawahara2001,Viswanath2007,Cvitanovic2010,Chandler2013, Lucas2015,Yalniz2021}. 
In practice, this imposes a limit on how unstable a UPO can be, severly limiting search as $Re$ increases \citep{Chandler2013}. A second issue is the strategy for recognising when the flow has nearly recurred. Typically this is done simply with an Euclidean norm of the difference between initial and final states, and consequently the threshold for near recurrence has to be set quite high. 
Finally, the third shortcoming is that many of the dynamically relevant periodic orbits are undiscovered, particularly those with a larger than average dissipation rate \citep{Chandler2013,Lucas2015}. These difficulties  have motivated a number of alternative approaches to guess generation \cite{Lan2004,Lucas2015,Farazmand2016, Page2020, Page2021, Parker2022} or even UPO identification \cite{LucasYasuda2022} but none have been sufficiently transformational to demonstrate the connection between dynamics and statistics even in the simplest model problems of steady turbulence. 

Even armed with a complete set of UPOs, a profound theoretical question is to predict the weights for how each UPO should be counted in a `UPO expansion' of the turbulent flow.  Periodic orbit theory \citep{Artuso1990a, Artuso1990b, Cvitanovic1991,ChaosBook} gives a theoretical prescription that is effective in low dimensional dynamical systems\citep[e.g.][]{Christiansen1997}. Yet, applying this theory to the Navier-Stokes equations remains challenging. An early attempt to apply this theory to the 2D Navier Stokes equations showed no greater skill than a control experiment of using equal weighting, although the set of solutions available was clearly too small \cite{Chandler2013, Lucas2015}. At this point, even determining whether an expansion of an arbitrary turbulent flow in terms of UPOs is an open question in fluid mechanics. 

In this paper we present fundamentally new approaches to both aspects of the problem which overcome many of the earlier limitations, introducing new methods for both finding and converging UPOs and for defining weights by labelling turbulent data according to which solution is closest in state space. 
In contrast to earlier work, our new method for UPO detection does not require careful construction of an initial guess, and yields large numbers of dynamically relevant UPOs.  
In order to do this, we adapt a recently developed fully differentiable flow solver \citep{Kochkov2021}, which allows us to find high-quality guesses for UPOs by performing gradient descent on a loss function involving entire solution trajectories.  This allows explicit searches for periodic orbits with certain properties (e.g. high dissipation rates) and can successfully converge large numbers of UPOs starting from arbitrary turbulent snapshots.  
We then train deep neural networks to learn accurate low order representations of the turbulence, which we are able to use to measure which UPO a turbulent orbit is closest to at any instant in time. 
The result is a Markovian turbulent dynamics, which not only allows us to define a set of weights for the UPOs via the chain's invariant measure, but also yields new insight into routes to extreme events.
The weights discovered from the invariant measure generate robust reproduction of the statistics of the turbulent attractor, including the full dissipation PDF, realising Hopf's original picture.
Although we develop these methods in a model problem, the underlying methodology promises to change our understanding of canonical flows at much higher $Re$.

\section{Periodic orbit search strategy}
\subsection{Two-dimensional turbulence}
\label{sec:flow_setup}
We demonstrate our new UPO search methodology within the widely studied turbulent `Kolmogorov' flow,
 monochromatically forced, two-dimensional turbulence in a doubly-periodic domain.
The governing equations are
\begin{subequations}
\begin{align}
    \partial_t \mathbf u + \mathbf u \cdot \boldsymbol \nabla \mathbf u &= -\boldsymbol \nabla p + \frac{1}{Re}\Delta \mathbf u + \sin (n y)\hat{\mathbf x}, \\
    \boldsymbol \nabla \cdot \mathbf u &= 0.
\end{align}
\label{eqn:ns_vel}
\end{subequations}
where the Reynolds number is defined as $Re := \sqrt{\chi}(L_y / 2\pi)^{3/2}/\nu$, with $\chi$ the forcing amplitude and $L_y$ the height of the computational domain. 
Throughout we set $L_x=L_y$ and use the popular forcing wavelength $n=4$ \cite[e.g.][]{Platt1991, Chandler2013, Farazmand2016, Lucas2015, Parker2022, LucasYasuda2022}. 
We also frequently use the out-of-plane vorticity, $\omega := (\boldsymbol \nabla \times \mathbf u)\cdot \hat{\mathbf z}$, both in the formulation for converging exact solutions (see appendix \ref{sec:app_comp}) and for training the neural networks used to label the solutions we find.
We consider two Reynolds numbers ($Re=40$ and $Re=100$) where self-sustaining turbulence is observed at both, with $Re=100$ clearly in the asymptotic regime \cite{Chandler2013}. 
At $Re=40$ around $50$ UPOs have been found previously, all with low average dissipation rates, while only $9$ UPOs have been converged at $Re=100$  \citep{Chandler2013}.

The governing equations (\ref{eqn:ns_vel}) are equivariant under continuous shifts in the horizontal direction, 
$$\mathscr T^{\alpha}: [u,v,\omega](x,y) \to [u,v,\omega](x+\alpha, y)$$
and consequently many of the simple invariant solutions are relative equilibria (travelling waves) or periodic orbits.
There are also discrete shift-reflect ($\mathscr S: [u,v,\omega](x,y) \to [-u,v,-\omega](-x, y+\pi/n)$, with $\mathscr S^{2n}\omega = \omega$) and rotation ($\mathscr R: [u,v,\omega](x,y) \to [-u,-v,\omega](-x, -y)$) symmetries. 
%
Here we seek relative unstable periodic orbits (RPOs) with some period $T$ and shift $\alpha$ which satisfy
\begin{equation}
    \mathscr T^{\alpha} \mathbf f^T (\mathbf u) - \mathbf u = \mathbf 0,
    \label{eqn:newton_standard}
\end{equation}
where $\mathbf f^t$ is the time-forward map of equation (\ref{eqn:ns_vel}). 

\subsection{Automatic differentiation for periodic orbits}
\begin{figure*}
    \centering
    \includegraphics[width=0.709\textwidth]{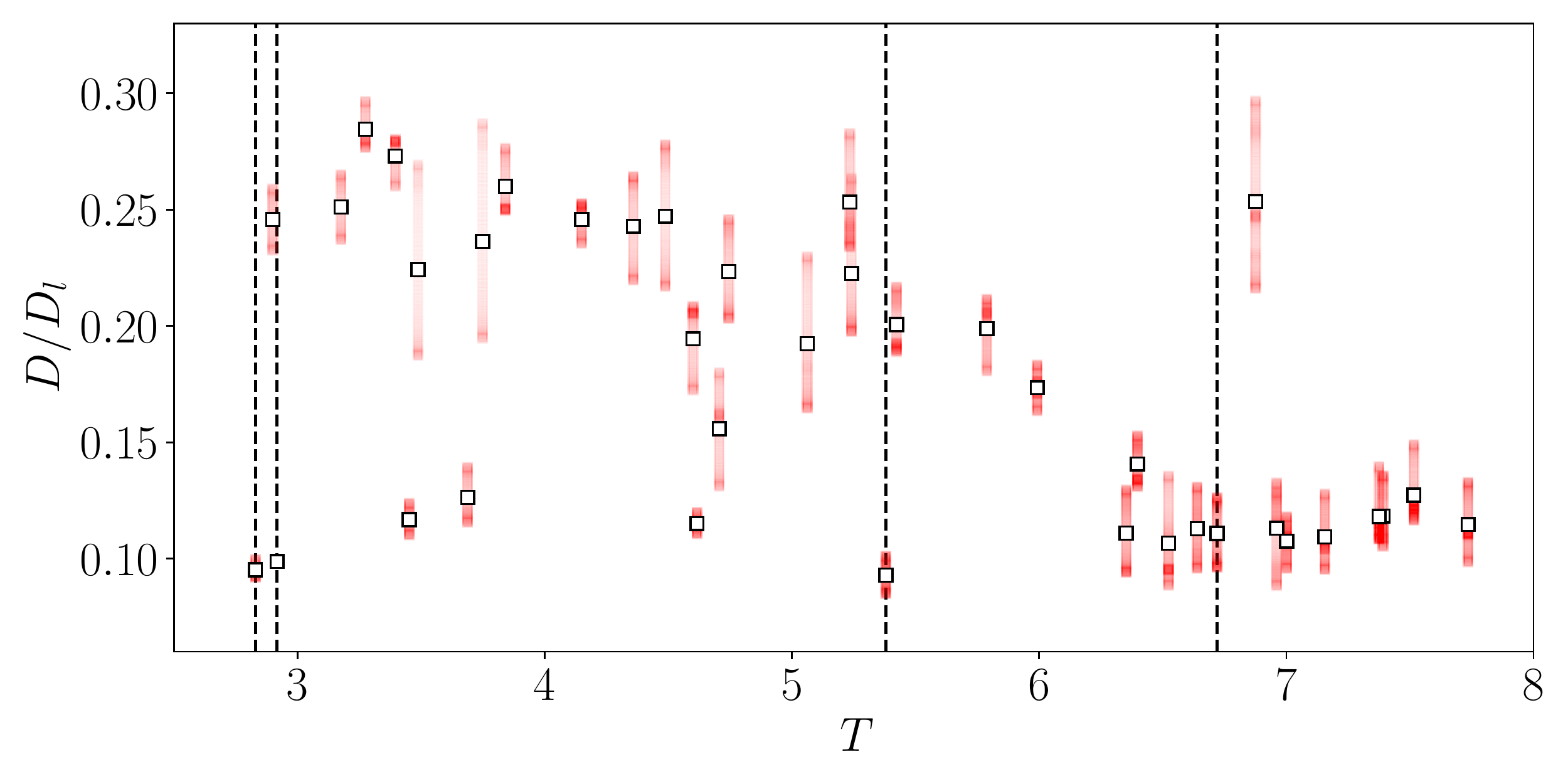}
    \includegraphics[width=0.281\textwidth]{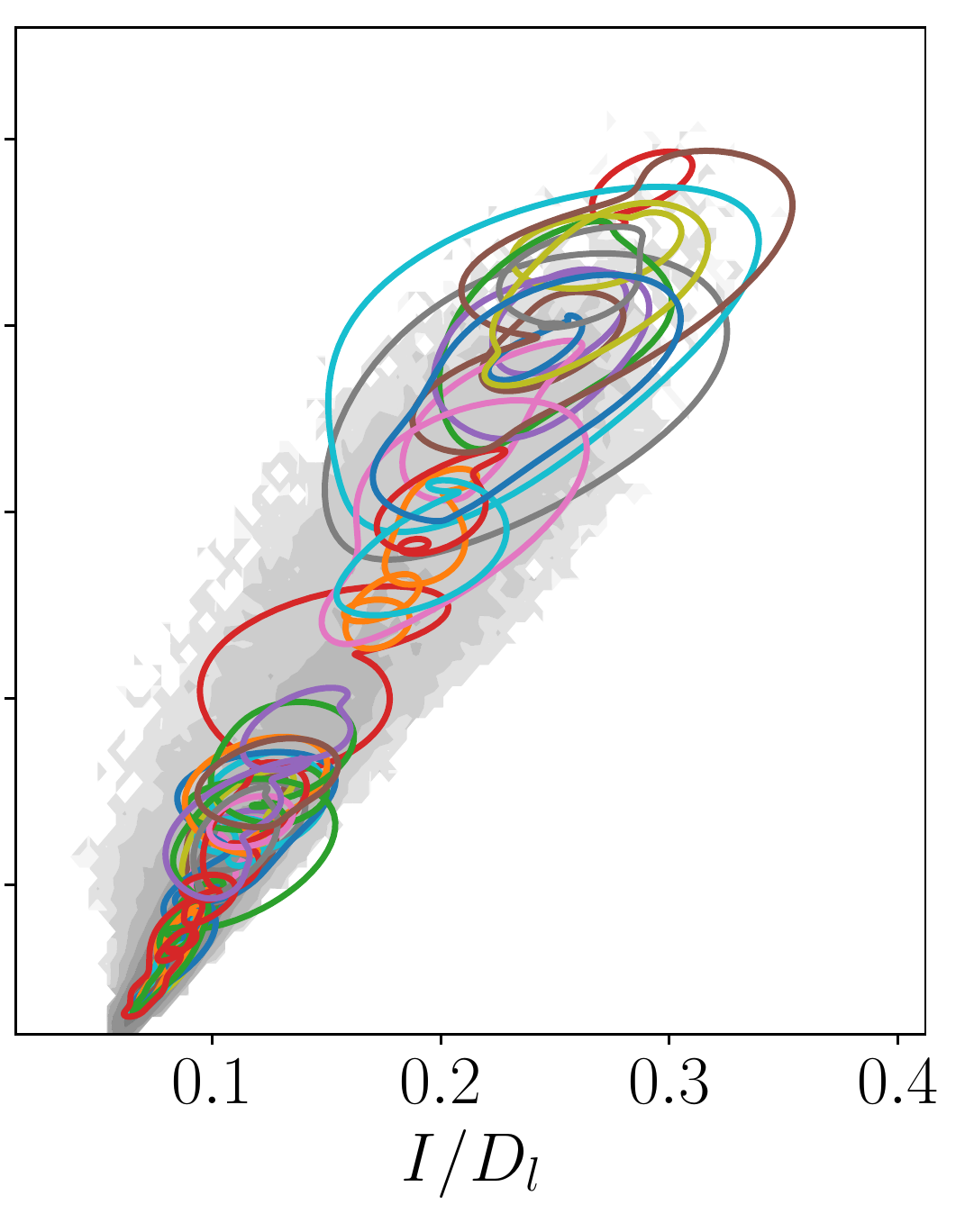}
    \caption{(Left) Dissipation against period of all converged orbits with periods $T<8$ at $Re=40$ (average dissipation rate is shown with a white square). The four previously known solutions are identified with vertical dashed lines at $T\in\{2.83, 2.92, 5.38, 6.72\}$. (Right) Energy production versus dissipation for all converged solutions at $Re=40$, including some longer orbits not shown on the left (all converged UPOs are listed in appendix \ref{sec:app_upo}). The probability density for the turbulent state (computed from a long calculation run for $t_l = 2.5 \times 10^5$) is shown in grey. Contour levels are spaced logarithmically with a minimum value of $10^{-6}$. All values are normalised by the laminar value $D_l = Re / (2n^2)$.}
    \label{fig:Re40_density}
\end{figure*}
We solve equation (\ref{eqn:ns_vel}) (and the equivalent velocity-vorticity form described in appendix \ref{sec:app_comp}) 
using \texttt{JAX-CFD}, a fully differentiable flow solver where gradients of the time-forward map, $\mathbf f^T(\mathbf u)$, with respect to initial conditions can be computed via automatic differentiation to machine precision \citep{Kochkov2021}. 
This capability forms the basis of the new periodic orbit search strategy.
We make use of both the `standard', finite-difference, primitive variable formulation \citep{Kochkov2021} and the spectral, vorticity version \citep{LCspectral} of \texttt{JAX-CFD}. 
The former is used to construct robust periodic orbit guesses, for reasons discussed below, and the latter for final convergence in a Newton solver and comparison to previously reported results \citep[which were all obtained in spectral codes][]{Chandler2013,Parker2022}. 

In contrast to earlier attempts to find UPOs by identifying `near recurrences' in time series and directly inputting them into a root-finder \citep{Kawahara2001,Viswanath2007,Cvitanovic2010,Chandler2013}, 
we instead conduct a search via gradient-based optimisation of a scalar loss function -- without any explicit initial condition selection.
This loss function is just a scaled norm of equation (\ref{eqn:newton_standard}),
\begin{equation}
    \mathscr L := \frac{\|\mathscr T^{\alpha}\mathbf f^T(\mathbf u) - \mathbf u\|}{\|\mathbf u\|},
    \label{eqn:loss_standard}
\end{equation}
and depends on the initial condition, $\mathbf u$, an unknown shift, $\alpha$ and period, $T$. 
Gradients with respect to all of these variables can be computed efficiently using the JAX library \citep{jax2018github} and its extensions \citep{kidger2021equinox}. 
Typically we deem that guesses for which we can reduce the loss to $\mathscr L \leq 0.015$ are suitable for passing to the Newton solve -- direct convergence with the optimiser is too slow and so it is used as an effective preconditioner on guesses for Newton.

At $Re=40$ we will also explicitly target certain periods, $T^*$, and attempt to find periodic orbits with average dissipation above some thresholds $D^*$ \citep[such UPOs were largely missing from previous results][]{Chandler2013,Lucas2015,Parker2022}. We do this by adding appropriate terms to the loss, 
\begin{subequations}
\begin{align}
    \mathscr L_T &:= \frac{\|\mathscr T^{\alpha}\mathbf f^T(\mathbf u) - \mathbf u\|}{\|\mathbf u\|} + \gamma \, (T - T^*)^2, \label{eqn:loss_T} \\
    \mathscr L_D &:= \frac{\|\mathscr T^{\alpha}\mathbf f^T(\mathbf u) - \mathbf u\|}{\|\mathbf u\|} + \kappa \, \sigma \left(\frac{D^* - \langle D \rangle_T}{\delta}\right), \label{eqn:loss_D}
\end{align}
\end{subequations}
where the hyperparameters $\gamma = 10^{-2}$, $\kappa = 100$ and $\delta = 10^{-2}$; $\sigma(\bullet)$ denotes a sigmoid function leading to very harsh penalisation if the dissipation time average falls below the threshold $D^*$. 
The penalisation on the target periods is relatively weak, with $|T-T^*| = O(1)$ resulting in a $O(10^{-2})$ contribution to the loss. 
When explicitly searching for high dissipation events we relax our threshold on Newton-worthy guesses to $\mathscr L_D=0.05$. 
We use an AdaGrad \citep{adagrad} optimiser with initial learning rate $\eta = 0.35$ throughout. 

The primitive variable formulation of \texttt{jax-cfd} \citep{Kochkov2021} allows for a \emph{constant} background vertical velocity $v_0$. 
The basic `Kolmogorov' flow described in \S\ref{sec:flow_setup} has $v_0=0$, and the addition of a finite $v_0 \neq 0$ fundamentally changes the system under consideration.
However, the addition of this effect 
prevents the optimiser from getting stuck in shallow local minima -- which is a common feature when searching directly with the spectral, vorticity formulation of \texttt{JAX-CFD}. 
Similar observations have been made in a recent attempt to find periodic orbits in a variational formulation, where non-solenoidal velocity fields were used as initial guesses \citep{Parker2022}. 
No constant background flow is possible in the spectral version of \texttt{jax-cfd} which is used to perform Newton convergence on the UPOs after the gradient-based optimisation, because we solve only for the rotational induced velocity field at each timestep. 

For short periods (e.g. $T \lesssim 10$ at $Re=40$), the weak background vertical velocity only weakly affects the UPO and the spectral Newton solve is capable of converging the nearby $v_0=0$ solution in a few steps. 
For longer periods the weak vertical flow has more of an impact, but can often successfully be efficiently removed with an additional optimisation run penalising the vertical velocity,
\begin{equation}
    \mathscr L_V := \frac{\|\mathscr T^{\alpha}\mathbf f^T(\mathbf u) - \mathbf u\|}{\|\mathbf u\|} + \mu \, v_0^2, 
\end{equation}
where we set $\mu = 10^3$ and use a smaller learning rate (typically $\eta = 10^{-2}$) to carefully deform the near-closed loop into one with near-zero vertical velocity.

\section{Unstable periodic orbits}
\subsection{Density of states at $Re=40$}

We first demonstrate the power of our new approach in the more well-studied problem of Kolmogorov flow at $Re=40$. 
This configuration has been examined by a number of previous authors \citep{Chandler2013,Lucas2015,Farazmand2016, Page2021,Parker2022} though we still know very little about the density of states as a function of their period, $\rho(T)$, and have not managed to identify any localised high dissipation UPOs. 
Motivated by both this and the importance of `prime cycles' in periodic orbit theory \citep{ChaosBook}, we conduct an intensive sweep over short periods $T\in [2, 10]$ using the period-targeting loss function (\ref{eqn:loss_T}). 
We increment the target $T^*$ in values of $0.5$ within this range and seed $50$ optimisation calculations at each $T^*$.
Each calculation is initialised with a random snapshot from the turbulent attractor, an initial guess for the period $T^0=T^*$ and zero initial shift, $\alpha^0=0$. 
Previously four solutions were known to exist in this range, with periods $T\in \{2.83, 2.92, 5.38, 6.72\}$. 

We also initialise separate searches for high dissipation solutions using loss function (\ref{eqn:loss_D}). 
We perform three computations searching for solutions with average dissipation rates above threshold values $D^* \in \{0.12, 0.15, 0.2\}$. 
The computations are initialised in a similar way to those described previously, though with the starting snapshots constrained to have dissipation values $D > D^*$. We fix the initial period in this search at $T^0=4$ and the shift again at $\alpha^0 = 0$. 

Large numbers of solutions are converged within the variable $v_0$ formulation, though success is not uniform across target $T^*$. 
For example, we find many solutions when $T^*\in \{5, 5.5, 6\}$, which when introduced into the $v_0=0$ spectral solver all collapse to the well known UPO at $T=5.38$ (similar behaviour is found close to $T=2.83$ and its integer multiples). 
Finding these common solutions is expected, but we do also find very large numbers of new solutions which were previously unknown. 
Indeed, for $T < 8$ we converge 38 unique solutions (detailed in appendix \ref{sec:app_upo}), including many high dissipation states that have been inaccessible to previous search methods. 

\begin{figure}
    \centering
    \includegraphics[width=0.49\textwidth]{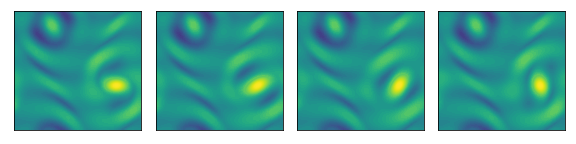}
    \includegraphics[width=0.49\textwidth]{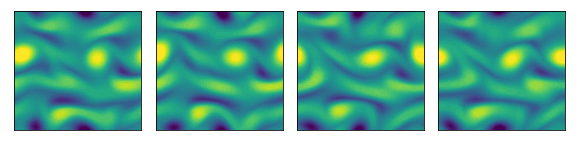}
    \includegraphics[width=0.49\textwidth]{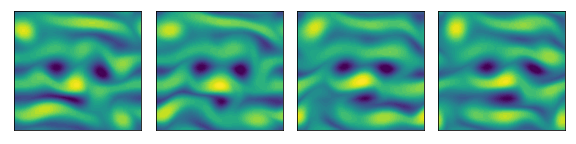}
    \caption{
    Spanwise vorticity are extracted at four points equispaced-in-time over three UPOs at $Re=40$.
    (Top) Shortest orbit at $Re=40$ -- the previously known $T=2.83$ solution with average dissipation rate $\langle D/D_l \rangle = 0.095$. 
    (Middle) A new high dissipation UPO with $T=2.90$ and average dissipation rate $\langle D / D_l \rangle = 0.246$.
    (Bottom) A new high dissipation UPO with $T=3.27$ and average dissipation rate $\langle D / D_l \rangle = 0.285$.
    In all cases contours run for $-10$ to $10$.}
    \label{fig:Re40_solutions}
\end{figure}
Remarkably, these short-period solutions appear to span nearly the full range of production and dissipation events in the overall flow (see the right panel in figure \ref{fig:Re40_solutions}). 
What is missing is actually the low dissipation events -- these are associated with slower dynamics and the UPOs tend to have longer periods.
To find these states, we also searched for solutions with target periods $T^*\in \{12.5, 15, 17.5, 20\}$. 
Generally, the optimiser struggles for these longer orbits, as $\mathbf f^T(\mathbf u)$ is likely far away on the solution manifold when starting from random initial conditions and gradient information is not particularly helpful. 
However, we do obtain several longer solutions -- most are listed in \cite{Chandler2013} -- which are all low dissipation. 
This aspect of the work would benefit from a more considered approach to initial condition selection to ensure that $\mathbf f^T(\mathbf u)$ is `nearby' on the solution manifold without necessarily requiring a near recurrence. 

We examine some of the $Re=40$ UPOs in figure \ref{fig:Re40_solutions}, where we report snapshots of spanwise vorticity at four points along the orbit. 
The low dissipation solutions (the $T=2.83$ UPO is shown in this figure) all share a common structure, with vortical structures sitting on top of a pair of slanted stripes of vorticity \citep[which are reminiscent of the first non-trivial equilbrium solution in this configuration, see][]{Page2021}. 
In the $T=2.83$ case the period corresponds to a complete rotation of an elliptical vortex patch located to the right of the panel; the orbit with $T=5.38$ (not shown) is similar but involves the rotation of a like-signed vortex pair. 

In contrast, the high dissipation structures display a much greater variety of dynamics.
For instance, one of the two `bursting' UPOs reported in figure \ref{fig:Re40_solutions} features a crystal like structure with four large-amplitude vortex cores that are maintained over the period.
The other high dissipation solution in figure \ref{fig:Re40_solutions} features a strong dipole structure that makes its way through the domain from left to right. 
Unsurprisingly, many of the new solutions have higher-dimensional unstable manifolds than those that have been previously documented (see full details in appendix \ref{sec:app_upo}).

\subsection{Periodic orbits at $Re=100$}
\begin{figure}
    \centering
    \includegraphics[width=0.45\textwidth]{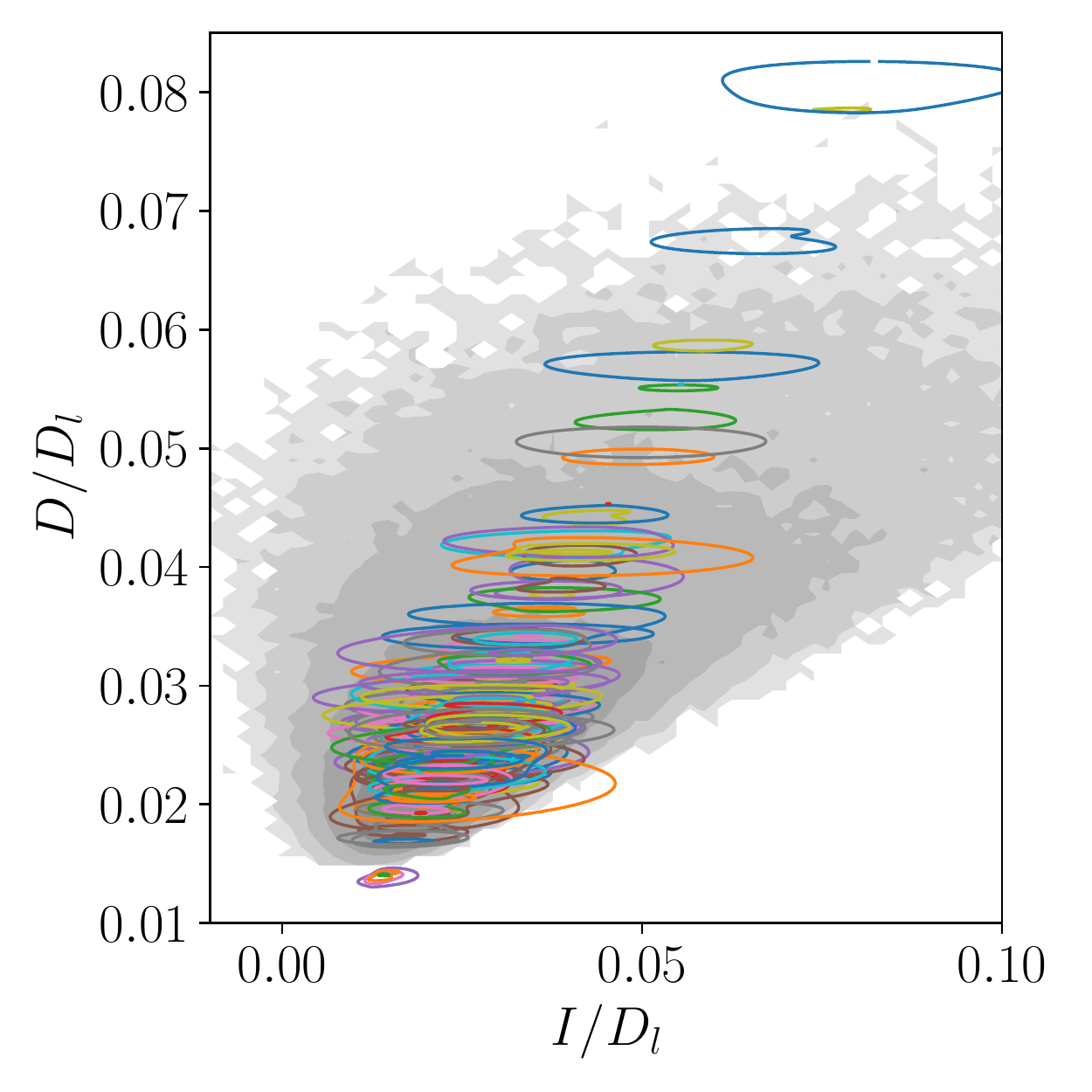}
    \caption{Energy production rate against dissipation rate at $Re=100$. Grey background is the PDF computed from a long turbulent computation with $T=2.5\times 10^5$. Contour levels are spaced logarithmically with a minimum value of $10^{-6}$. Closed loops are the two-dimensional projections of the 151 converged UPOs. All values are normalised by the laminar value $D_l = Re / (2n^2)$. All periodic orbits along with relevant properties (period, shift, Floquet exponents) are listed in appendix \ref{sec:app_upo}.}
    \label{fig:Re100_ID}
\end{figure}
We now turn our attention to the strongly turbulent case at $Re=100$, where only a handful of solutions have been obtained by any previous method \citep{Chandler2013}. 
The 9 solutions obtained previously at this value of $Re$ were all of short period, $T < 5$. 
The majority of solutions we report here were computed using the basic loss function (\ref{eqn:loss_standard}) without additional physics. 

We initialise large numbers of computations with starting periods selected from $T^0\in \{1, 2, 3, 4, 5\}$, with additional calculations also performed $T^0=2.5$, motivated by the success rate of the $T^0 = 2$ and $T^0=3$ calculations.
We stopped these computations once a batch of $O(100)$ guesses failed to yield a new solution not already contained in our collection of converged UPOs. 
We also perform a high dissipation search using loss function (\ref{eqn:loss_D}) for periodic orbits with average dissipations $\langle D /D_l\rangle \geq 0.03$, though find this to be much less effective than the equivalent computations at $Re=40$. 
In contrast, the standard loss function (\ref{eqn:loss_standard}) returns a wide range of periodic orbits without augmentation, rather than the same subset of solutions. 
Overall, we observe a success rate for converging solutions of between $5-15$\%, depending on the choice of starting period. 
It is important to emphasise that the starting point in the process is a \emph{random} snapshot from a turbulent calculation, and should be contrasted to other methods with a success rate close to zero where guesses were more carefully constructed \citep[e.g. the recurrent flow analysis in][produced 9 solutions from an analysis of $10^5$ time units of data]{Chandler2013}.

\begin{figure}
    \centering
    \includegraphics[width=0.49\textwidth]{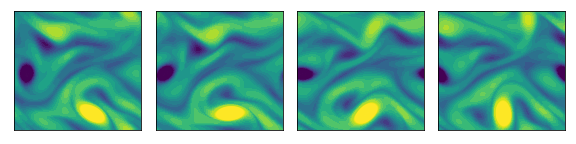}
    \includegraphics[width=0.49\textwidth]{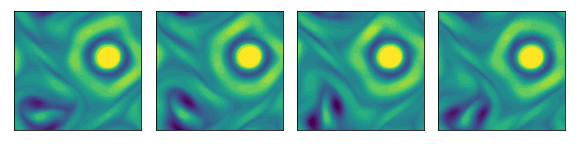}
    \includegraphics[width=0.49\textwidth]{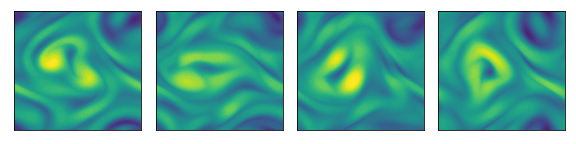}
    \includegraphics[width=0.49\textwidth]{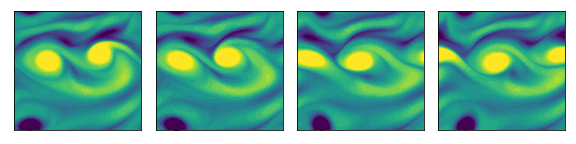}
    \caption{Spanwise vorticity are extracted at four points equispaced-in-time over four UPOs at $Re=100$. From top-to-bottom the UPOs have the following periods and average dissipation rates: $(T, \langle D / D_l \rangle) = (1.424, 0.057)$, $(1.794, 0.053)$, $(4.212, 0.027)$ and $(1.164, 0.078)$ (for full details of converged solutions see appendix \ref{sec:app_upo}). Vorticity contour levels run between $\pm 10$.}
    \label{fig:Re100_solutions}
\end{figure}
Our computations have so far yielded 151 unique UPOs at $Re=100$, and the states are summarised in figure \ref{fig:Re100_ID} (and listed in full in appendix \ref{sec:app_upo}). 
Here, we see that most of the UPOs appear to be highly localised in phase space, nearly all appearing as small closed loops in the two-dimensional projection. 
The states appear even more localised when visualised in terms of their kinetic energy, $E$ (not shown), and 
full coverage of the associated turbulent PDFs will require many more states than the equivalent computations at $Re=40$.
%
%

We visualise some of the converged UPOs in figure \ref{fig:Re100_solutions}. 
The solutions show a wide variety of different dynamical behaviours. 
Often, the states are dominated by two large vortex patches \citep[expected due to the inverse cascade][]{Boffetta_ARFM} -- see middle two rows of figure \ref{fig:Re100_solutions}.
There is wide variation within this type of solution, in addition to the behaviour in figure (e.g. where we see a strong stationary vortex and a co-rotating pair) there are also states with three or more like-signed co-rotating vortices -- see further details in appendix \ref{sec:app_upo}. 
The more strongly dissipative states (top and bottom rows of figure \ref{fig:Re100_solutions}) show an even greater variety of vorticity dynamics, and further work will be required to assess the various `classes' of UPO discovered by our new approach.

\section{Markov chains from periodic orbit shadowing}
\label{sec:markov}
\subsection{Identifying UPO shadowing}
Given the wide range of converged UPOs found, and the particularly good coverage of the $I-D$ plane at $Re=40$, we now attempt to label snapshots of a turbulent time series by which UPO is `closest', in an attempt to verify and visualise the original conjecture by \citet{Hopf1948}. 
Our objective is to use our library of UPOs to partition the state space and convert turbulent orbits to realisations of a discrete-time Markov process.

To accurately measure distance to the nearest UPO we have trained highly-accurate deep convolutional autoencoders in a `DenseNet' \citep{huang2017densely,huang2019convolutional} configuration (see appendix \ref{sec:app_comp} for full architectural and training details), and will use an observable based on the latent representations in these networks, rather than using a distance between snapshots in physical space. 
The accuracy of these models has been demonstrated in our recent work \citep{PHBK23} over a wide range of $Re$ -- accuracy is maintained even on the rare, highest dissipation events. 

The autoencoders consist of an encoder, $\mathscr E: \mathbb R^{N_x\times N_y} \to \mathbb R^m$ (here $m=128$ at $Re=40$ and $m=512$ at $Re=100$), and decoder $\mathscr D : \mathbb R^m \to  \mathbb R^{N_x\times N_y}$ such that $[\mathscr D \circ \mathscr E](\omega) \approx \omega$.
Given an encoded snapshot, $\mathscr E(\omega)$, we first construct a streamwise-shift invariant observable by projecting $\mathscr E$ onto so-called `latent Fourier modes', which are eigenvectors of a discrete shift operator $\mathbf T_{\alpha} \mathscr E(\omega) := \mathscr E(\mathscr T^{\alpha}\omega)$ for some fixed streamwise shift $\alpha$. 
We then use these projections to build a vector observable $\boldsymbol{\psi}(\omega)$, which has the property $\boldsymbol \psi(\mathscr T^s \omega) = \boldsymbol \psi(\omega)$ $\forall s \in \mathbb R$  (full details in appendix \ref{sec:app_comp}).
Finally, we compute the period-averaged value of $\boldsymbol \psi$ for each periodic orbit as well as each of its 15 discrete-symmetry copies, $\{\langle \boldsymbol{\psi}(\mathscr S^m \mathscr R^q f^t(\omega_j))\rangle_T \, : \, 0 \leq m \leq 7, \, q \in\{0, 1\}\}$.

The nearest periodic orbit to a snapshot $\omega$ is then determined according to 
\begin{equation}
    j^* = {\arg \, \min}_j \, \text{min}_{m, q}\| \boldsymbol{\psi}(\omega) - \langle \boldsymbol{\psi}(\mathscr S^m \mathscr R^q f^t(\omega_j))\rangle_T\|_2,
    \label{eqn:po_label}
\end{equation}
where $1 \leq j \leq N_p$ ($N_p$ is the total number of periodic orbits in our library at a given $Re$) and we search over the discrete symmetries at every time instant.
A comparison to the time-average of the UPO embeddings is robust as the majority of our UPOs are short and localised in phase space, though more sophisticated methods could search over the time direction too. 

We construct long trajectories of length $T=2.5 \times 10^4$ at $Re=40$ and $T=10^4$ at $Re=100$, where snapshots are separated by $\delta t = 1$ in the former case and $\delta t = 0.25$ in the latter. 
Snapshot spacing in each is motivated by the typical period of the UPOs in our library (e.g. $T\sim 5$ is common at $Re=40$ while a plurality of solutions at $Re=100$ have $1\leq T\leq 2$) and the motivation to be able to observe `shadowing' of periodic solutions if such dynamics occurs. 
The turbulent time series are converted to sequences of labels of the form $\text{PO}_i \to \text{PO}_i \to \text{PO}_i \to \text{PO}_k \to \text{PO}_j \to \text{PO}_j \to \cdots$ (for example).

\subsection{Discrete-time Markov chains and statistical predictions}
\begin{figure*}
    \centering
    \includegraphics[width=\textwidth]{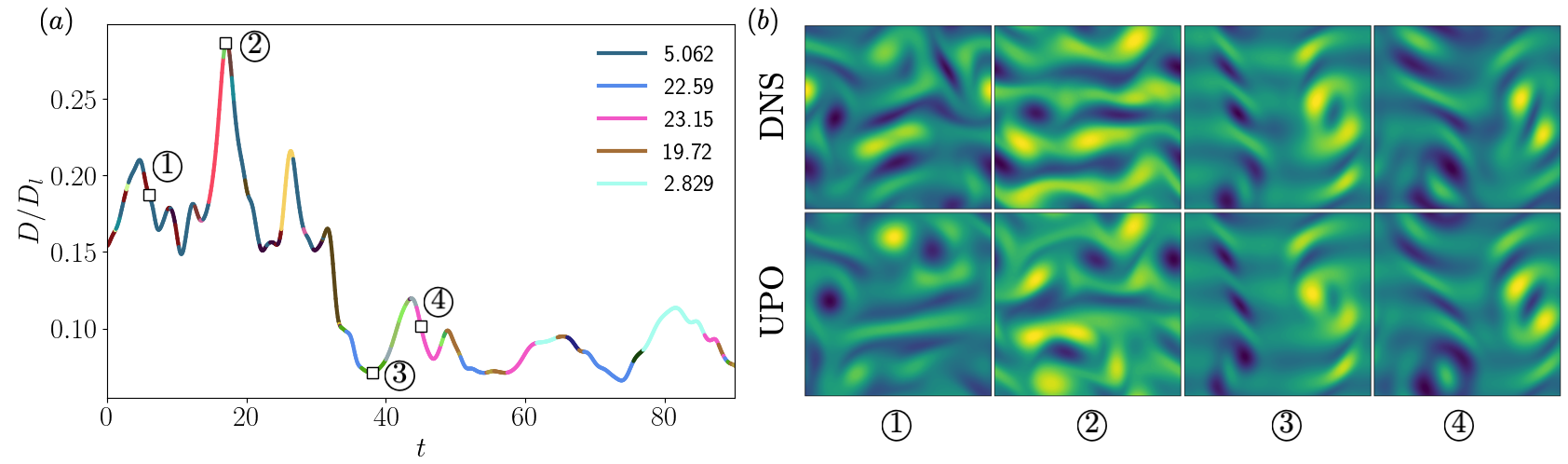}
    \caption{(a) Example time evolution of dissipation rate at $Re=40$, coloured according to the nearest UPO as determined by equation (\ref{eqn:po_label}). The five most frequently visited UPOs in this interval are highlighted in the legend, labelled by their period, $T$. (b) Snapshots of vorticity (top) from the direct numerical simulation (DNS) used to generate the dissipation plot at times identified with squares/numbers in (a), along with snapshots from the closest UPO (bottom), where we have selected the horizontal shift, $s$, and the time along the orbit, $\tau$, to minimise $\|\omega - \mathscr T^s f^{\tau}(\omega_{PO})\|_2$.}
    \label{fig:upo_traj_compare}
\end{figure*}
An example of the UPO-labelling protocol described above is reported in figure \ref{fig:upo_traj_compare}, where we have used a much finer $\delta t$ for illustrative purposes. 
The example in the figure shows an extended high-dissipation bursting event which returns to more quiescent low-dissipation dynamics at around $t\sim 30$. 
The curve is coloured according to which UPO is closest as determined by equation (\ref{eqn:po_label}), which indicates that there are extended periods of time where the flow remains close to a particular UPO, and that this particular sequence can be well described by just a small number of exact solutions. 
In fact, this example trajectory spends more than half its time in the vicinity of just four UPOs, with the high dissipation event repeatedly sampling the same solution (dark blue curves in figure \ref{fig:upo_traj_compare}, see also the figure legend). 
We compare snapshots along this example orbit with snapshots from the `closest' UPO in the right panel of the figure, and can see many of the same qualitative flow features from the turbulence reproduced in the periodic solutions. 
It is also clear that the match could be improved -- a more robust labelling protocol could also search over the time direction of all UPOs, albeit at significantly increased computational expense.

\begin{figure}
    \centering
    \includegraphics[width=0.49\textwidth]{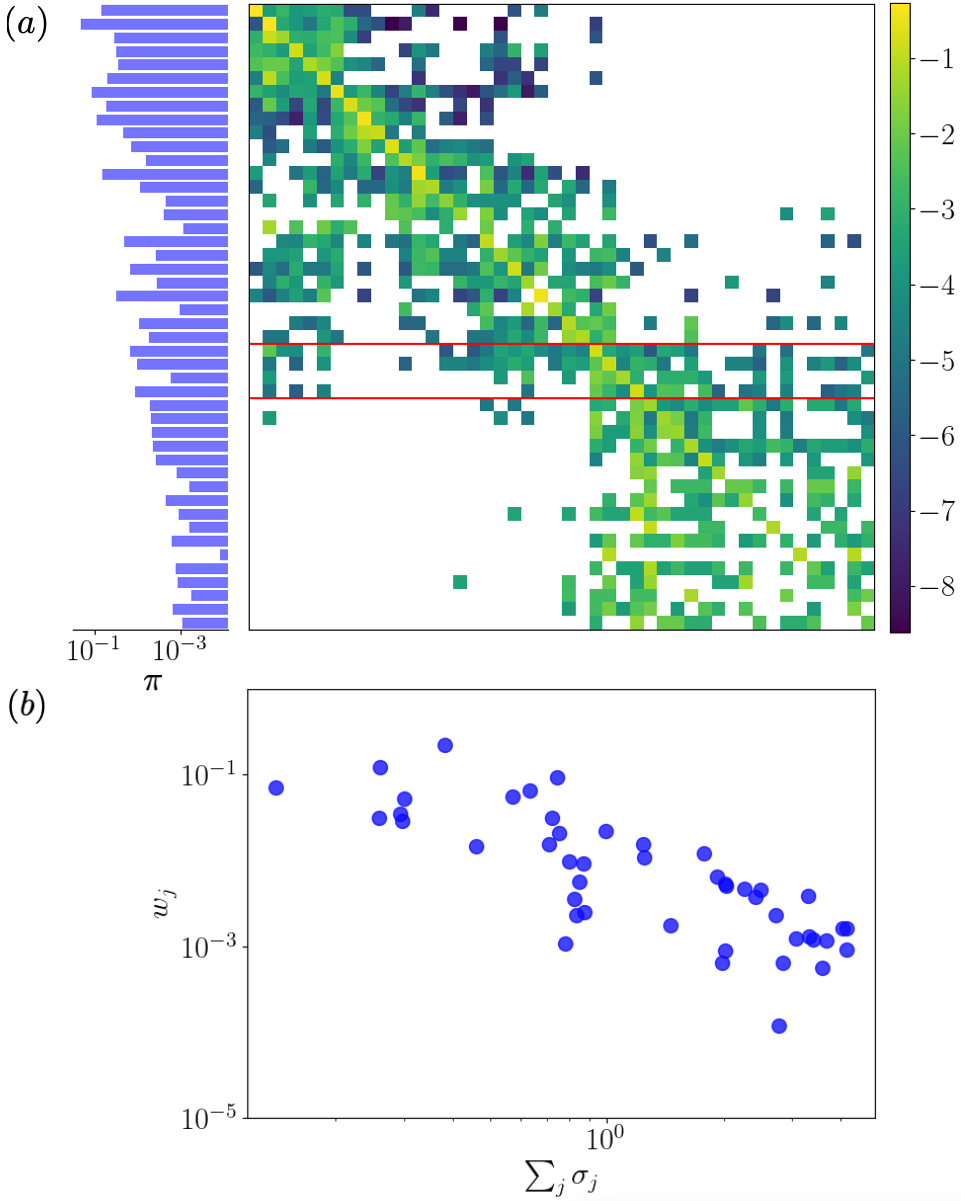}
    \caption{
    (a) Invariant measure $\boldsymbol \pi$ and transition matrix $\mathbf P$ at $Re=40$ (log of transition probabilities is shown, spacing between snapshots is $\delta t = 1$). States are ordered from lowest to highest average dissipation rate (lowest at top/leftmost). The red lines identify `mixing' states discussed in the text.  
    (b) The weights in the expansion (\ref{eqn:upo_pdf}) -- which are also the invariant measure of the Markov chain $w_j = \pi_j$ -- plotted against the (real part of the) sum of growing Floquet exponents $\sum_j \sigma_j$, $\sigma_j > 0$, for each UPO.
    }
    \label{fig:Re40_transitions}
\end{figure}
We now use the long time series described at the end of the previous section 
to construct transition probability matrices $\mathbf P$, with elements $P_{ij} := \mathbb P(\text{PO}_i \to \text{PO}_j)$.
This quantity is trivially computed by simple counting of the transitions between states before normalising rows $\sum_j P_{ij} = 1 \, \forall i$. 
The transition matrix at $Re=40$ is reported in figure \ref{fig:Re40_transitions}(a), where the states have been ordered by dissipation (low dissipation top rows, highest at the bottom). 
We also show the invariant measure obtained via $\boldsymbol \pi^T \mathbf P = \boldsymbol \pi^T$.

There are a number of interesting features present in the transition matrix which merit further discussion. 
The transition matrix generally shows the largest probabilities along its diagonal, which means that for most states the most likely outcome is to remain in the vicinity of that particular UPO for another time instant ($\delta t = 1$ here at $Re=40$). This is consistent with a turbulent orbit shadowing individual recurrent solutions. 
The off-diagonal non-zero probabilities also show that transitions tend to occur between states with similar dissipation rates. 
The chaotic dynamics at $Re=40$ can be delineated into a low-dissipation `quiescent' regime and rarer, high-dissipation `bursting' events \citep[roughly with normalised dissipation $D/D_l \gtrsim 0.15$][]{Chandler2013,Farazmand2016,Page2021}, and this delineation is clear in the transition matrix of figure \ref{fig:Re40_stats};
there are a multitude of routes between the high dissipation states, though transitions to these bursting events appear to occur via a small number $\sim 3$ of gateway, lower-dissipation UPOs. 

The Markovian view of turbulence can be extended to make statistical predictions in the spirit of periodic orbit theory \citep{Artuso1990a,Artuso1990b,ChaosBook}. 
To do this we seek a fixed set of weights, $\{w_j\}_{j=1}^{N_p}$, where $N_p$ is the total number of UPOs found, such that \emph{any} statistic $\Gamma$ can be be constructed as a linear superposition of the statistics of the UPOs,
\begin{equation}
    \Gamma(\mathbf w) = \sum_{j=1}^{N_p} w_j \Gamma_j.
    \label{eqn:upo_pdf}
\end{equation}
From our transition matrix, we are able to define the weights simply as $w_j \equiv \pi_j$, where $\sum_j w_j = 1$ by definition. 
These weights are examined as a function of the instability of the underlying UPOs in the lower panel (b) of figure \ref{fig:Re40_transitions} (as measured by the sum of the growth rates from Floquet exponents). The weights are anti-correlated with the level of instability, where the highly unstable states are much less important in the reconstruction.  
This reflects the fact that the highly dissipative UPOs (which contribute to the very weak right tail of the distribution) tend to be much more unstable.


\begin{figure*}
    \centering
    \includegraphics[width=\textwidth]{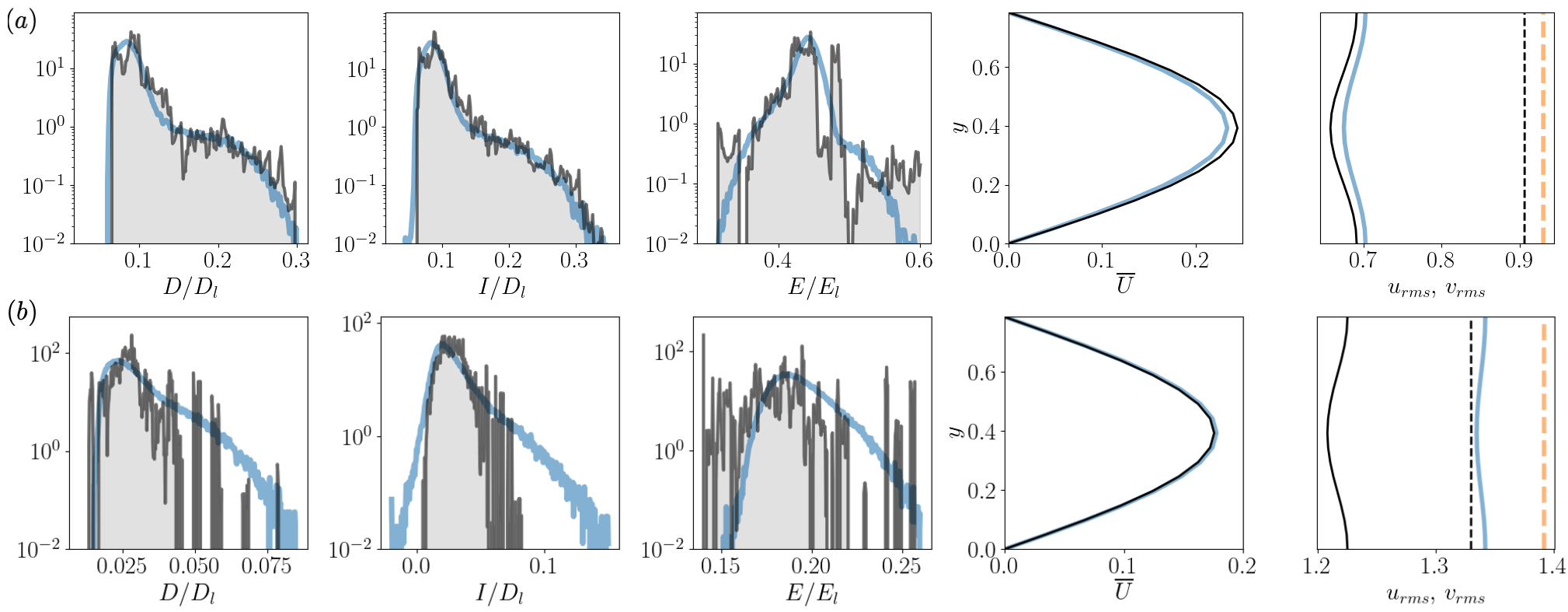}
    \caption{
    UPO-based predictions of statistics computed from the invariant measure of the Markov chain, which defines the weights in the UPO expansion (equation \ref{eqn:upo_pdf}), for (a) $Re=40$ and (b) $Re=100$ (transition matrix reported in appendix \ref{sec:app_Re100}). 
    First three panels from left to right show PDFs of dissipation rate, production rate and energy: the dashed blue lines are `ground truth' PDFs obtained from a long turbulent calculation ($T = 2.5\times 10^5$) and the filled grey curves are the UPO reconstructions.
    Final two panels compare the mean velocity profile $\overline{U}(y)$ and the root-mean-square velocity fluctuations ($u$ and $v$ are left and right in the panel) -- averaged over the streamwise direction, discrete symmetries and time. Blue and orange curves are the DNS `ground truth' for $u$ and $v$ respectively, black curves the UPO reconstruction.
    }
    \label{fig:all_stats}
\end{figure*}
To demonstrate the performance of the above UPO expansion (\ref{eqn:upo_pdf}), figure \ref{fig:all_stats}(a) reports PDFs of the dissipation rate, energy production rate and kinetic energy computed over a long time horizon ($T=2.5\times 10^5$) at $Re=40$ overlayed with UPO-based PDFs computed via expressions like (\ref{eqn:upo_pdf}).
The reproduction of both dissipation and production is remarkably complete -- although we are missing some of the very lowest values associated with the longer orbits (as discussed previously). 
Most notable in both cases is the accurate reconstruction of the high-$D$/high-$I$ tails, which has eluded all previous attempts \citep{Chandler2013,Lucas2014}.
The kinetic energy, $E$, is also reproduced to a fairly high standard \citep[in particular compare to past attempts to apply periodic orbit theory][]{Chandler2013}.
The missing states at $E/E_l \sim 0.35$ and $E/E_l \sim 0.45$ can be plausibly linked to the missing lowest dissipation orbits.

The UPO predictions (again with the same weights) for the mean velocity profile and root-mean-square velocity fluctuations are also tested in figure \ref{fig:all_stats}(a) \citep[averaged over discrete symmetries as well as the streamwise coordinate and time, see][]{Chandler2013,Farazmand2016}. 
All three are close to the true time-averaged values (e.g. errors $O(1 - 2\%)$ in $u_{rms}$ and $v_{rms}$),
which again is a dramatic improvement in UPO-based prediction over the past state-of-the-art. 

A similar analysis is performed using our large collection of 151 UPOs at $Re=100$ in figure \ref{fig:all_stats}(b).
The corresponding transition matrix is included in appendix \ref{sec:app_Re100} and again indicates shadowing but with likely transitions widely separated in average dissipation -- though clearly the picture is incomplete because we are missing many important states as implied by our earlier results (e.g. figure \ref{fig:Re100_ID}).
Nonetheless, the statistics reported in figure \ref{fig:all_stats}(b) are promising and represent a significant advance on what has been possible previously even at much lower $Re$. 
In particular, the mean profile is produced near perfectly, while the errors on the RMS velocities are only $O(10\%)$. 
The PDFs indicate that the missing states are associated with larger dissipation and energy values, and the UPO-search procedure outlined here provides a clear strategy for filling these gaps with further computation.

%
%
\section{Conclusion}
%
%

In this paper we have assembled the first compelling evidence in support of Hopf's early view of turbulence \citep{Hopf1948} as a high-dimensional pinball bouncing between unstable simple invariant sets. 
To do this we both (i) designed a new methodology for finding large numbers of dynamically-relevant UPOs -- a long-standing limitation in the field -- and 
(ii) presented an approach to accurately label turbulent snapshots by the `closest' UPO, where distance is measured in the latent space of a deep convolutional autoencoder. 
The result is a Markovian picture of turbulence, which yields both new insight into dynamical pathways and routes to extreme events along with robust statistical predictions for the chaotic dynamics. 

%
%
The new UPO search strategy is formed as a gradient-based optimisation problem and is implemented in a fully differentiable flow solver. 
The loss-based approach allows for a targeted search for solutions with specific features (e.g. high dissipation, high energy)
and application of the method at modest $Re=40$ revealed very large numbers of new solutions with short periods -- both high- and low-dissipation -- that had been previously undetectable. 
The apporach remains effective at the much higher value of $Re=100$, where
 we again found very large numbers of short solutions.
The states at high $Re$ appear to be highly localised in state space, and display a wealth of interesting vortical dynamics. 

To label vorticity snapshots according to the `closest' UPO, we trained highly-accurate deep convolutional autoencoders and measured similarity using an observable in the low-dimensional latent space of these networks. 
We were then able to treat long turbulent time series as Markov chains with each UPO being a distinct state, building transition matrices and then using their invariant measure to make statistical predictions. 
The approach was particularly effective at $Re=40$, where the new library of UPOs covers nearly the full range of production and dissipation events seen by the fully turbulent state and as a result statistical predictions are robust. 
Even with an incomplete set of states at $Re=100$, the statistical predictions are fairly robust and a substantial improvement on what has been possible even at much lower $Re$ using earlier methods. 

Despite the enormous numbers of new solutions converged, there are still important missing states at $Re=40$ and $Re=100$. 
The statistical gaps (e.g. the lower dissipation events at $Re=40$) could benefit from an improved process of selecting the snapshots which are input to the optimiser. 
In the present configuration, improvements could involve something as simple as only selecting snapshots which return to a similar region of state space -- not a near recurrence but a weaker requirement e.g. determined by the autoencoders trained here -- or by searching over discrete symmetries. 
This is likely an important consideration when studying more complex three-dimensional flows.


The new approach to UPO search outlined in this paper allows us to explicitly target the gaps in the PDFs we have constructed so far, and we believe that it is this ability to explicitly search for dynamics of interest, combined with the effectiveness of the loss-based approach at high-$Re$, that will make this an effective strategy in three-dimensional multiscale turbulence. 
Most importantly for future work, we have observed a steady improvement in the statistical results as we have converged more states over the production of this manuscript. 
All this suggests that a representation of turbulence with a UPO expansion {\em is} a viable modelling approach, which opens up exciting new vistas for prediction and control.

\appendix
\section{Additional computational details}
\label{sec:app_comp}
\subsection{Simulations}
Simulations with the finite-difference version of \texttt{JAX-CFD} were performed on grids of size $N_x \times N_y = 256 \times 256$ at $Re=40$ and $N_x \times N_y = 512$ at $Re=100$. 
For the Newton-solve component of our algorithm we use the spectral version of \texttt{JAC-CFD} \citep{LCspectral}, and we matched the resolution in the spectral simulations to those in the precursor finite difference optimisation. 
The timestep was determined by a CFL condition based on a velocity estimate which is twice that of the laminar base profile, $2 \times Re/(2 n^2)$, which is typically much larger than velocities observed in the turbulent regime. 

The spectral version of the code solves the Navier-Stokes equations in vorticity-velocity form,
\begin{equation}
    \partial_t \omega + \mathbf u \cdot \boldsymbol \nabla \omega = \frac{1}{Re}\Delta \omega - n\cos(ny),
\label{eqn:vorticity}
\end{equation}
with $\omega := (\boldsymbol \nabla \times \mathbf u)\cdot \hat{\mathbf z}$ (compare to equation \ref{eqn:ns_vel}). 
Unlike the finite-difference, primitive variable formulation, no background constant flow is possible.
The velocity field at each timestep is that induced by the vorticity, and is found from the solution of a Poisson equation, 
$\Delta \psi = -\omega$, where the streamfunction $\psi$ yields the induced velocity components via $u = \partial_y \psi$, $v= -\partial_x \psi$. 

\subsection{Neural network and distance metric}

\emph{Architecture:}
The convolutional autoencoders used in \S \ref{sec:markov} are a combination of an encoder, $\mathscr E_m$, and decoder, $\mathscr D_m$, such that
\begin{equation}
    \mathscr A_m (\omega) := [\mathscr D_m \circ \mathscr E_m](\omega) \approx \omega.
\end{equation}
Dimensionality reduction is performed with the encoder $\mathscr E_m : \mathbb R^{N_x\times N_y} \to \mathbb R^m$, where we fix $m=128$ for the $Re=40$ data and $m=512$ at $Re=100$. 
Performance at a variety of $m$ for various $Re$ is examined in \citet{PHBK23}. 

The encoder is a fully convolutional architecture, with dimensionality reduction done by max pooling after a `dense block' of convolutions \citep{huang2017densely,huang2019convolutional}.
Each dense block consists of three successive convolutional layers, with each receiving the output of the previous convolution concatenated with outputs of all other upstream layers within the same dense block. 
Each convolution within a dense block creates an additional 32 feature maps.
After each dense block we apply max pooling followed by a single convolutional layer to reduce the number of feature maps to $32$, and the full encoder is made up of six dense-block/max pooling combinations. 
At the inner-most representation, the encoder produces an image of shape $4 \times 8 \times M$, were $M = m / 32$ ($m$ is the specified dimensionality of the inner-most latent representation). As described in \citep{PHBK23}, the vertical value of `8' is set by the discrete shift reflect symmetry in the Kolmogorov flow. 

Throughout the network we use `GELU' activation functions \citep{gelu_arxiv}, apart from the decoder output where $\tanh$ is used (input data is normalised $\omega \to \omega / \omega_{\text{norm}}$ such that $\max |\omega(x, y)| \leq 1$). 
The decoder module is similar in structure to the encoder but in reverse, with upsampling applied in place of max pooling. 
Code and weights for the model will be released on publication and linked here, and further details can be found in \citet{PHBK23}. 

\emph{Training:}
We use a loss function which is a modified `mean square error':
\begin{equation}
    \mathscr L_{AE} := \frac{1}{2N}\sum_{j=1}^N \| \mathscr A_m (\omega_j) - \omega_j \|^2 + \frac{1}{2N}\sum_{j=1}^N\| \mathscr A_m^2 (\omega_j) - \omega_j^2 \|^2,
\end{equation}
where the additional term (a mean square error on the vorticity squared) is designed to encourage the network to learn an effective representation of the -much rarer- high dissipation events (rather than changing the distribution the underlying data is drawn from) where this term is an increasingly important contribution to the overall loss.

We train on $N=10^5$ samples generated at each $Re$, consisting of 1000 independent trajectories with snapshots spaced by an advective time unit. 
We apply data augmentation: randomly shifting in $x$ and $y$, as well as applying the rotational symmetry. 
An Adam optimizer \citep{Kingma2015} is used for all models, with learning rate $\eta = 5\times 10^{-4}$, and we train for $500$ epochs with a batch size of 64. 
The performance of these models is examined in detail in \citep{PHBK23} and the reader is referred there for further detail.

\emph{Latent Fourier analysis:}
Latent Fourier analysis is an interpretability technique for autoencoders originally described in \citep{Page2021} which exploits the continuous symmetry in the governing equations/boundary conditions. 
In latent Fourier analysis we seek an operator to perform continuous shifts for embeddings of vorticity fields, i.e. 
\begin{equation}
    \mathbf T_{\alpha} \mathscr E_m(\omega) := \mathscr E(\mathscr T^{\alpha}\omega),
\end{equation}
where we pick $\alpha = 2\pi /n$ with $n\in \mathbb N$, such that $\mathbf T_{\alpha}^n \mathscr E(\omega) = \mathscr E(\omega)$.
The eigenvalues of $n$ are then the $n^{th}$ roots of unity, $\lambda_j = \exp(2\pi i l_j / n)$, and we refer to $l_j\in \mathbb Z$ as a latent wavenumber \citep{Page2021,PHBK23}.

In practice, we determine an approximate $\widehat{\mathbf T}_{\alpha}$ using the `dynamic mode decomposition' algorithm of \citet{Schmid2010}, which gives us the best (in a least squares sense) operator that maps between the test dataset of embeddings and their $\alpha$-shifted counterparts. 
Empirically we observe a small number of non-zero latent wavenumbers which saturate as $\alpha$ is incrementally reduced ($l_{max} = 3$ at $Re=40$ and $l_{max} = 7$ at $Re=100$) which indicates that the networks embed patterns with a fundamental periodicity set by $l$. Each latent wavenumber is then (potentially highly) degenerate.

We can then write down a representation of the embedding of a snapshot subject to an arbitrary shift $s\in \mathbb R$:
\begin{equation}
    \mathscr E_m(\mathscr T^s \omega) =  \sum_l \mathscr P^l(\mathscr E_m(\omega))e^{ils},
\end{equation}
which makes a clear connection to a standard Fourier transform. Here $\mathscr P^l$ is the projector onto the degenerate eigenspace associated with wavenumber $l$:
\begin{equation*}
    \mathscr P^l(\mathscr E_m(\omega)) = \sum_{k=1}^{d(l)}\mathscr P^{l}_k(\mathscr E_m(\omega)),
\end{equation*}
and the projectors onto each of the $d(l)$ degenerate directions are defined as $\mathscr P^{l}_k(\mathscr E_m(\omega)):= [(\boldsymbol{\xi}^{(l)\dagger}_k)^H\mathscr E_m(\omega)]\boldsymbol{\xi}^{(l)}_k$, where the $\boldsymbol \xi_k^{(l)}$ are the eigenvectors in subspace $l$, while the dagger indicates the adjoint eigenvectors.

To define our shift invariant observable $\boldsymbol \psi(\omega)$ used in \S \ref{sec:markov} we perform an SVD of projections onto individual eigenspaces associated with $l\leq 3$ \citep[the analysis of][indicates that the majority of energy is contained in these patterns]{PHBK23}, determining the degeneracy $d(l)$ by the number of eigenvalues $|\lambda| > 0.9$. 
The shift invariant observable is then determined using the left singular vectors from the SVD within each subspace, $\{\mathbf u_l^k\}_{k=1}^{d(l)}$:
\begin{equation}
        \boldsymbol \psi(\omega) := \begin{pmatrix}
        (\mathbf u_{l=0}^1)^H \mathscr P^{0}(\mathscr E(\omega)) \\
        (\mathbf u_{0}^2)^H \mathscr P^{0}(\mathscr E(\omega)) \\
        \vdots \\
        |(\mathbf u_{1}^1)^H \mathscr P^{1}(\mathscr E(\omega))| \\
        \vdots \\
        |(\mathbf u_{3}^{d(3)})^H \mathscr P^{3}(\mathscr E(\omega))|
    \end{pmatrix}.
\end{equation}
Taking the absolute value of the projections onto $l>0$ ensures that $\psi(\omega) \approx \psi(\mathscr T^{s}\omega)\, \forall s \in \mathbb R$.

The utility and relevance of this observable to features in the flow, as compared to e.g. measuring distances between vorticity fields in physical space, was established in \citet{PHBK23}, where it has been shown (i) that individual projections onto latent Fourier modes can be decoded (using the decoder $\mathscr D_m$) into physically meaningful patterns resembling known simple invariant solutions and (ii) measuring similarity between snapshots using $\psi$ over $\omega$ as the observable significantly improves recurrent flow analysis for UPO detection.

\section{Further results at $Re=100$}
\label{sec:app_Re100}
\begin{figure}
    \centering
    \includegraphics[width=0.49\textwidth]{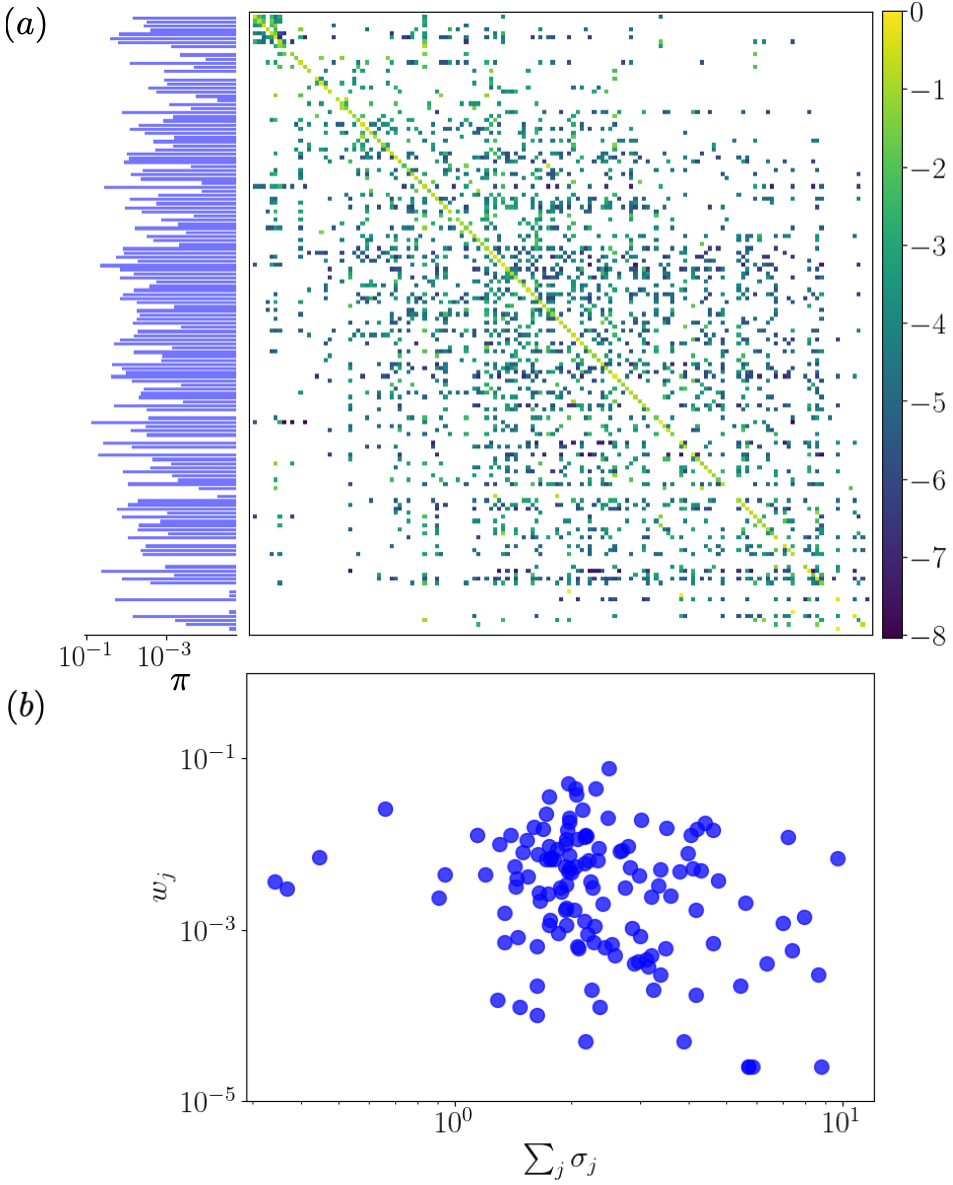}
    \caption{
    (a) Invariant measure $\boldsymbol \pi$ and transition matrix $\mathbf P$ at $Re=100$ (log of transition probabilities is shown, spacing between snapshots is $\delta t = 0.25$). States are ordered from lowest to highest average dissipation rate (lowest at top/leftmost).  
    (b) The weights in the expansion (\ref{eqn:upo_pdf}) -- which are also the invariant measure of the Markov chain $w_j = \pi_j$ -- plotted against the (real part of the) sum of growing Floquet exponents $\sum_j \sigma_j$, $\sigma_j > 0$, for each UPO.
    }
    \label{fig:Re100_transitions}
\end{figure}
The higher $Re=100$ transition matrix is included here in figure \ref{fig:Re100_transitions} (compare to the $Re=40$ results in figure \ref{fig:Re40_transitions}), along with the invariant measure used to compute the statistics in figure \ref{fig:all_stats}. 
There is no clear distinction between low/high dissipation states and transitions can apparently occur between widely separated UPOs (in terms of dissipation).
However, these results are clearly likely to be impacted significantly by the large number of missing states -- e.g. see the $I-D$ plot in figure \ref{fig:Re100_ID}, and a clearer picture will likely emerge as we continue to converge new solutions in future calculations.
The relationship between the weights defined by the invariant measure of the $Re=100$ transition matrix and the unstable growth rates of the associated UPOs are also reported in figure \ref{fig:Re100_transitions}. 
Unlike the $Re=40$ results, it is challenging to identify a clear relationship, which again may be resolved after computation of more solutions. 

\begin{figure}
    \centering
    \includegraphics[width=0.49\textwidth]{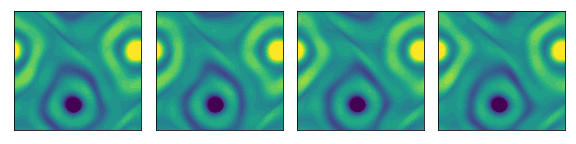}
    \includegraphics[width=0.49\textwidth]{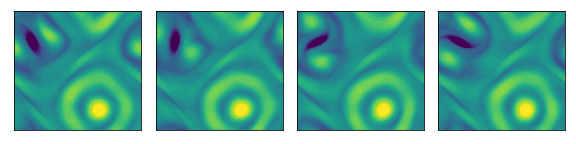}
    \includegraphics[width=0.49\textwidth]{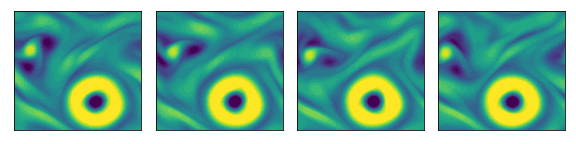}
    \includegraphics[width=0.49\textwidth]{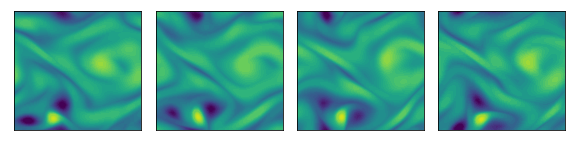}
    \includegraphics[width=0.49\textwidth]{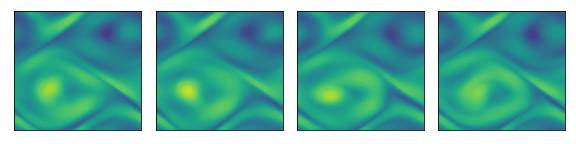}
    \caption{Spanwise vorticity are extracted at four points equispaced-in-time over five UPOs at $Re=100$. From top-to-bottom the UPOs have the following periods and average dissipation rates: $(T, \langle D / D_l \rangle) = (1.356, 0.038)$, $(1.723, 0.021)$, $(1.966, 0.068)$, $(2.196, 0.027)$ and $(2.590, 0.02)$ (for full details of converged solutions see table \ref{table:periodic_orbits_Re100}). Vorticity contour levels run between $\pm 10$.}
    \label{fig:Re100_additional}
\end{figure}
In addition to the transition matrix, we also report snapshots of spanwise vorticity for five further UPOs at $Re=100$ in figure \ref{fig:Re100_additional} to demonstrate the wealth of vorticity dynamics contained in the UPO library. 
These include localised co-rotating three-vortex states, as well as large quiescent vortex patches.

\section{Periodic orbit details}
\label{sec:app_upo}
Here we report details of the UPOs we have found via automatic differentiation at both $Re=40$ and $Re=100$ in tables \ref{table:periodic_orbits_Re40} and \ref{table:periodic_orbits_Re100} respectively, including their leading Floquet exponent and the dimension of the unstable manifold. 

\begin{table}
\caption{Periodic orbits at $Re=40$. Known solutions as listed in \citet{Chandler2013} are listed in the `UPO' column; $T$ is the period, $\alpha$ the shift. $N$ indicates the number of unstable directions and $\sigma_r$ is the growth rate in the leading Floquet exponent. We also report the average dissipation normalised by the laminar value, $\langle D/D_l \rangle$.}
\label{table:periodic_orbits_Re40}
\begin{tabular*}{8cm}{P{1cm}|P{1.25cm}|P{1.25cm}|P{1cm}|P{1.25cm}|P{1cm}}
\hline
UPO & $T$ & $\alpha$ & $N$ & $\sigma_r$ & $\langle D / D_l\rangle$ \\
\hline
  R19 & 12.207 & 0.243 & 2 & 0.07 & 0.072 \\
   & 22.597 & 0.257 & 4 & 0.209 & 0.081 \\
  R24 & 19.779 & 0.248 & 6 & 0.202 & 0.081 \\
  R6 & 20.808 & 0.06 & 3 & 0.172 & 0.083 \\
  R22 & 19.723 & 0.222 & 4 & 0.172 & 0.083 \\
   & 18.771 & 0.457 & 3 & 0.105 & 0.084 \\
  R34 & 23.157 & 0.265 & 3 & 0.113 & 0.084 \\
  P1 & 5.38 & 0.0 & 7 & 0.191 & 0.093 \\
  P2 & 2.83 & 0.0 & 5 & 0.223 & 0.095 \\
  P3 & 2.917 & 0.0 & 7 & 0.236 & 0.099 \\
   & 6.523 & 0.0 & 6 & 0.201 & 0.107 \\
   & 7.002 & 0.194 & 6 & 1.376 & 0.108 \\
   & 7.156 & 0.032 & 4 & 0.292 & 0.109 \\
  R4 & 6.72 & 0.106 & 8 & 0.343 & 0.111 \\
   & 6.352 & 0.156 & 6 & 0.34 & 0.111 \\
   & 6.639 & 0.151 & 7 & 0.34 & 0.113 \\
   & 6.961 & 0.0 & 5 & 1.386 & 0.113 \\
   & 7.735 & 0.101 & 7 & 0.28 & 0.115 \\
   & 4.616 & 0.545 & 12 & 2.081 & 0.115 \\
   & 8.067 & 0.742 & 7 & 0.243 & 0.116 \\
   & 3.452 & 0.362 & 11 & 0.247 & 0.117 \\
   & 7.375 & 0.0 & 7 & 0.301 & 0.118 \\
   & 7.391 & 0.081 & 8 & 0.219 & 0.118 \\
   & 3.689 & 0.392 & 8 & 0.258 & 0.126 \\
   & 7.516 & 0.193 & 9 & 0.24 & 0.127 \\
   & 6.397 & 0.439 & 11 & 0.19 & 0.141 \\
   & 4.706 & 0.173 & 10 & 0.253 & 0.156 \\
   & 5.992 & 0.099 & 12 & 0.274 & 0.173 \\
   & 5.062 & 1.324 & 10 & 0.253 & 0.192 \\
   & 4.6 & 0.615 & 10 & 0.341 & 0.194 \\
   & 5.789 & 0.236 & 12 & 0.394 & 0.199 \\
   & 5.423 & 0.372 & 13 & 0.348 & 0.201 \\
   & 5.241 & 1.206 & 11 & 0.366 & 0.223 \\
   & 4.745 & 0.842 & 13 & 0.378 & 0.223 \\
   & 3.488 & 1.142 & 12 & 0.444 & 0.224 \\
   & 3.749 & 0.0 & 11 & 0.393 & 0.236 \\
   & 4.359 & 0.288 & 14 & 0.366 & 0.243 \\
   & 2.901 & 0.486 & 13 & 0.356 & 0.246 \\
   & 4.15 & 0.384 & 15 & 0.581 & 0.246 \\
   & 4.488 & 1.28 & 20 & 0.526 & 0.247 \\
   & 3.176 & 0.424 & 15 & 0.504 & 0.251 \\
   & 5.234 & 0.526 & 15 & 0.559 & 0.253 \\
   & 6.876 & 0.188 & 15 & 0.534 & 0.254 \\
   & 3.841 & 0.945 & 14 & 0.497 & 0.26 \\
   & 3.396 & 0.926 & 18 & 0.479 & 0.273 \\
   & 3.275 & 0.588 & 17 & 0.462 & 0.285 \\
\hline
\end{tabular*}
\end{table}

\begin{table}
\caption{Periodic orbits at $Re=100$. Known solutions as listed in \citet{Chandler2013} are listed in the `UPO' column; $T$ is the period, $\alpha$ the shift. $N$ indicates the number of unstable directions and $\sigma_r$ is the growth rate in the leading Floquet exponent. We also report the average dissipation normalised by the laminar value, $\langle D/D_l \rangle$.}
\label{table:periodic_orbits_Re100}
\begin{tabular*}{8cm}{P{1cm}|P{1.25cm}|P{1.25cm}|P{1cm}|P{1.25cm}|P{1cm}}
\hline
UPO & $T$ & $\alpha$ & $N$ & $\sigma_r$ & $\langle D / D_l\rangle$ \\
\hline
   & 4.654 & 0.073 & 9 & 0.098 & 0.014 \\
  R14 & 4.526 & 0.071 & 9 & 0.105 & 0.014 \\
   & 4.58 & 0.072 & 7 & 0.093 & 0.014 \\
   & 4.607 & 0.071 & 8 & 0.089 & 0.014 \\
   & 4.524 & 0.023 & 9 & 0.386 & 0.017 \\
   & 3.34 & 0.048 & 10 & 2.614 & 0.017 \\
   & 6.741 & 0.104 & 12 & 0.13 & 0.018 \\
   & 2.986 & 0.525 & 12 & 0.294 & 0.019 \\
   & 2.976 & 0.532 & 15 & 0.315 & 0.019 \\
   & 2.756 & 0.075 & 13 & 0.527 & 0.02 \\
   & 2.285 & 0.435 & 16 & 0.359 & 0.02 \\
   & 2.695 & 0.076 & 8 & 0.48 & 0.02 \\
   & 6.569 & 0.216 & 11 & 1.35 & 0.02 \\
   & 2.231 & 0.427 & 19 & 0.365 & 0.02 \\
   & 2.241 & 0.077 & 11 & 0.411 & 0.02 \\
   & 2.053 & 0.072 & 12 & 0.563 & 0.02 \\
   & 2.59 & 0.067 & 13 & 0.539 & 0.02 \\
   & 2.594 & 0.045 & 15 & 0.491 & 0.02 \\
   & 2.362 & 0.102 & 10 & 0.368 & 0.021 \\
   & 2.362 & 0.465 & 14 & 0.392 & 0.021 \\
   & 3.943 & 0.06 & 17 & 2.243 & 0.021 \\
   & 4.361 & 0.199 & 12 & 0.599 & 0.021 \\
   & 2.441 & 0.002 & 10 & 0.347 & 0.021 \\
   & 2.442 & 0.002 & 10 & 3.981 & 0.021 \\
   & 2.281 & 0.438 & 12 & 0.306 & 0.021 \\
   & 2.398 & 0.12 & 10 & 0.472 & 0.021 \\
   & 2.601 & 0.057 & 18 & 0.259 & 0.022 \\
   & 4.45 & 0.246 & 10 & 0.503 & 0.022 \\
   & 2.098 & 0.118 & 9 & 0.297 & 0.022 \\
   & 4.669 & 0.169 & 9 & 0.297 & 0.022 \\
   & 2.33 & 0.138 & 13 & 0.378 & 0.022 \\
   & 2.465 & 0.052 & 17 & 3.682 & 0.022 \\
   & 2.281 & 0.087 & 9 & 0.416 & 0.022 \\
   & 4.488 & 0.152 & 10 & 0.462 & 0.022 \\
   & 5.793 & 0.065 & 14 & 0.218 & 0.022 \\
   & 2.527 & 0.092 & 12 & 0.403 & 0.022 \\
   & 2.742 & 0.492 & 11 & 0.444 & 0.022 \\
   & 2.018 & 0.553 & 10 & 0.399 & 0.022 \\
   & 2.176 & 0.117 & 14 & 0.269 & 0.022 \\
   & 1.709 & 0.043 & 13 & 0.429 & 0.022 \\
  R16 & 1.938 & 0.121 & 6 & 0.271 & 0.023 \\
   & 2.29 & 0.013 & 13 & 0.394 & 0.023 \\
   & 3.536 & 0.651 & 15 & 0.254 & 0.023 \\
   & 4.052 & 0.215 & 18 & 0.586 & 0.023 \\
   & 2.301 & 0.037 & 10 & 0.429 & 0.023 \\
   & 2.777 & 0.48 & 17 & 0.408 & 0.023 \\
   & 3.84 & 0.078 & 15 & 2.336 & 0.023 \\
   & 3.573 & 0.363 & 14 & 0.272 & 0.024 \\
   & 4.422 & 0.443 & 15 & 0.262 & 0.024 \\
   & 1.944 & 0.545 & 14 & 0.452 & 0.024 \\
   & 4.635 & 0.386 & 15 & 0.26 & 0.024 \\
   & 2.61 & 0.089 & 16 & 0.504 & 0.024 \\
\hline
\end{tabular*}
\end{table}

\setcounter{table}{1} 
\begin{table}
\caption{Continued}
\begin{tabular*}{8cm}{P{1cm}|P{1.25cm}|P{1.25cm}|P{1cm}|P{1.25cm}|P{1cm}}
\hline
UPO & $T$ & $\alpha$ & $N$ & $\sigma_r$ & $\langle D / D_l\rangle$ \\
\hline
   & 3.312 & 0.874 & 13 & 0.419 & 0.024 \\
   & 2.662 & 0.12 & 15 & 0.542 & 0.024 \\
   & 2.592 & 0.069 & 13 & 0.481 & 0.024 \\
   & 2.82 & 0.046 & 14 & 0.385 & 0.025 \\
   & 1.984 & 0.122 & 17 & 0.556 & 0.024 \\
   & 4.321 & 0.93 & 14 & 0.436 & 0.025 \\
   & 2.2 & 0.012 & 13 & 0.454 & 0.025 \\
   & 5.821 & 0.089 & 14 & 0.318 & 0.025 \\
   & 4.497 & 0.109 & 15 & 0.368 & 0.025 \\
   & 3.231 & 0.12 & 11 & 0.663 & 0.025 \\
   & 4.457 & 0.644 & 12 & 0.293 & 0.025 \\
   & 3.359 & 0.329 & 12 & 0.381 & 0.025 \\
   & 3.611 & 0.802 & 12 & 2.539 & 0.025 \\
   & 4.464 & 0.079 & 12 & 0.408 & 0.025 \\
   & 3.736 & 0.338 & 9 & 0.546 & 0.026 \\
   & 3.93 & 0.347 & 11 & 0.398 & 0.026 \\
   & 3.303 & 0.385 & 16 & 0.423 & 0.026 \\
   & 3.829 & 0.378 & 13 & 2.325 & 0.026 \\
   & 6.32 & 0.588 & 10 & 0.263 & 0.026 \\
   & 1.899 & 0.299 & 15 & 0.323 & 0.026 \\
   & 3.919 & 0.941 & 15 & 0.394 & 0.026 \\
   & 3.588 & 0.577 & 11 & 0.259 & 0.026 \\
   & 5.175 & 0.138 & 13 & 0.616 & 0.026 \\
   & 3.869 & 0.148 & 13 & 0.23 & 0.026 \\
   & 2.205 & 0.032 & 14 & 4.383 & 0.026 \\
   & 3.186 & 0.836 & 15 & 0.43 & 0.026 \\
   & 4.139 & 0.396 & 10 & 0.496 & 0.026 \\
   & 2.398 & 0.399 & 10 & 0.279 & 0.026 \\
   & 4.099 & 0.533 & 14 & 0.312 & 0.026 \\
   & 3.736 & 0.992 & 16 & 2.526 & 0.026 \\
   & 3.685 & 1.039 & 12 & 0.418 & 0.027 \\
   & 2.163 & 0.382 & 18 & 4.197 & 0.027 \\
   & 2.196 & 0.166 & 13 & 0.432 & 0.027 \\
   & 4.09 & 0.027 & 15 & 0.324 & 0.027 \\
   & 3.581 & 0.169 & 9 & 0.243 & 0.027 \\
   & 4.162 & 0.756 & 15 & 0.255 & 0.027 \\
   & 1.917 & 0.113 & 16 & 0.575 & 0.027 \\
   & 2.238 & 0.346 & 11 & 0.328 & 0.027 \\
  R17  & 3.827 & 0.008 & 16 & 0.818 & 0.027 \\
   & 2.136 & 0.025 & 11 & 0.422 & 0.027 \\
   & 1.881 & 0.195 & 9 & 0.275 & 0.027 \\
   & 3.506 & 0.094 & 17 & 2.639 & 0.027 \\
   & 3.993 & 0.292 & 15 & 0.505 & 0.027 \\
   & 4.212 & 0.765 & 13 & 0.335 & 0.027 \\
   & 3.228 & 0.407 & 15 & 0.406 & 0.028 \\
   & 2.688 & 0.18 & 21 & 0.476 & 0.028 \\
   & 3.248 & 0.909 & 13 & 0.481 & 0.028 \\
   & 1.345 & 0.007 & 7 & 1.113 & 0.028 \\
   & 3.757 & 0.17 & 11 & 0.22 & 0.028 \\
   & 3.51 & 0.068 & 13 & 2.621 & 0.028 \\
   & 3.311 & 0.885 & 11 & 0.417 & 0.028 \\
   & 1.328 & 0.09 & 24 & 0.656 & 0.029 \\
   & 1.999 & 0.049 & 14 & 0.643 & 0.029 \\
   & 4.0 & 0.003 & 10 & 0.32 & 0.029 \\
   & 1.178 & 0.081 & 22 & 0.671 & 0.029 \\
   & 2.123 & 0.319 & 11 & 0.349 & 0.029 \\
   & 3.645 & 0.169 & 15 & 0.249 & 0.03 \\
\hline
\end{tabular*}
\end{table}

\setcounter{table}{1} 
\begin{table}
\caption{Continued}
\begin{tabular*}{8cm}{P{1cm}|P{1.25cm}|P{1.25cm}|P{1cm}|P{1.25cm}|P{1cm}}
\hline
UPO & $T$ & $\alpha$ & $N$ & $\sigma_r$ & $\langle D / D_l\rangle$ \\
\hline
   & 1.711 & 0.207 & 8 & 0.267 & 0.03 \\
   & 2.135 & 0.036 & 16 & 0.376 & 0.03 \\
   & 1.941 & 0.425 & 14 & 4.804 & 0.031 \\
   & 1.861 & 0.241 & 18 & 0.414 & 0.031 \\
   & 1.723 & 0.204 & 15 & 0.525 & 0.031 \\
   & 3.63 & 0.561 & 15 & 0.364 & 0.031 \\
   & 1.729 & 0.206 & 14 & 0.245 & 0.031 \\
   & 1.313 & 0.022 & 24 & 0.815 & 0.032 \\
   & 3.65 & 0.037 & 15 & 0.992 & 0.032 \\
   & 3.774 & 0.577 & 16 & 0.597 & 0.032 \\
   & 3.498 & 0.077 & 11 & 0.306 & 0.033 \\
   & 3.623 & 0.063 & 10 & 2.572 & 0.033 \\
   & 2.031 & 0.168 & 10 & 0.37 & 0.034 \\
   & 1.246 & 0.02 & 9 & 1.167 & 0.034 \\
   & 1.744 & 0.014 & 17 & 5.067 & 0.034 \\
   & 1.757 & 0.028 & 14 & 1.175 & 0.034 \\
   & 3.438 & 0.184 & 15 & 2.733 & 0.035 \\
   & 1.788 & 0.085 & 11 & 1.034 & 0.036 \\
   & 1.71 & 0.404 & 12 & 0.403 & 0.037 \\
   & 1.356 & 0.126 & 19 & 0.896 & 0.038 \\
   & 1.601 & 0.051 & 18 & 0.505 & 0.038 \\
   & 1.718 & 0.376 & 14 & 0.483 & 0.039 \\
   & 3.113 & 0.44 & 20 & 0.827 & 0.039 \\
   & 2.877 & 0.056 & 26 & 1.229 & 0.04 \\
   & 1.993 & 0.032 & 23 & 1.285 & 0.041 \\
   & 1.635 & 0.363 & 14 & 0.438 & 0.041 \\
   & 1.763 & 0.111 & 11 & 0.439 & 0.041 \\
   & 1.746 & 0.027 & 20 & 0.468 & 0.042 \\
   & 1.858 & 0.378 & 14 & 0.509 & 0.042 \\
   & 1.821 & 0.082 & 13 & 1.142 & 0.044 \\
   & 2.005 & 0.028 & 15 & 1.402 & 0.045 \\
   & 1.883 & 0.228 & 19 & 0.958 & 0.045 \\
   & 1.287 & 0.163 & 19 & 1.037 & 0.049 \\
   & 1.425 & 0.846 & 12 & 0.629 & 0.05 \\
   & 1.794 & 0.012 & 16 & 1.582 & 0.053 \\
   & 1.946 & 0.0 & 16 & 1.615 & 0.055 \\
   & 2.658 & 0.336 & 14 & 2.863 & 0.055 \\
   & 1.424 & 0.815 & 12 & 6.785 & 0.057 \\
 P4 & 1.185 & 0.0 & 16 & 1.201 & 0.059 \\
   & 1.966 & 0.435 & 20 & 1.536 & 0.068 \\
   & 1.164 & 0.768 & 14 & 1.062 & 0.078 \\
   & 1.939 & 0.0 & 29 & 1.724 & 0.08 \\
\hline
\end{tabular*}
\end{table}

%


\end{document}